\documentclass[prb,twocolumn,showpacs,superscriptaddress]{revtex4-1}

\usepackage{amsmath}
\usepackage{amssymb}
\usepackage{xspace}
\usepackage{graphicx}
\usepackage{grffile}
\usepackage{nicefrac}
\usepackage{psfrag}
\usepackage{color}

\graphicspath{{figs/}}

\begin{document}

\title{Formation of orbital-selective electron states in 
LaTiO$_3$/SrTiO$_3$ superlattices}

\author{Frank Lechermann}
\affiliation{I. Institut f{\"u}r Theoretische Physik,
Universit{\"a}t Hamburg, D-20355 Hamburg, Germany}
\author{Lewin Boehnke}
\affiliation{I. Institut f{\"u}r Theoretische Physik,
Universit{\"a}t Hamburg, D-20355 Hamburg, Germany}
\author{Daniel Grieger}
\affiliation{International School for Advanced Studies (SISSA),
and CNR-IOM Democritos, Via Bonomea 265, I-34136 Trieste, Italy}

\begin{abstract}
The interface electronic structure of correlated LaTiO$_3$/SrTiO$_3$ superlattices
is investigated by means of the charge self-consistent combination of the local density 
approximation (LDA) to density functional theory (DFT) with dynamical mean-field theory 
(DMFT). Utilizing a pseudopotential technique together with a continuous-time quantum 
Monte-Carlo approach, the resulting complex multiorbital electronic states are addressed 
in a coherent fashion beyond static mean-field. General structural relaxations are taken 
into account on the LDA level and cooperate with the driving forces from strong 
electronic correlations. This alliance leads to an Ti($3d_{xy}$) dominated low-energy 
quasiparticle peak and a lower Hubbard band in line with photoemission studies. 
Furthermore correlation effects close to the band-insulating bulk SrTiO$_3$ limit as 
well as the Mott-insulating bulk LaTiO$_3$ limit are studied via realistic single-layer 
embeddings.

\end{abstract}

\pacs{73.20.-r, 71.27.+a, 71.15.-m}
\maketitle

The research on layered heterostructures composed of different metal oxide
(MO) compounds emerges as a major new field in condensed matter physics.~\cite{hwa12} 
Especially the intriguing appearance of a two-dimensional (2D) electron gas  
from interfacing bulk-insulating MOs may open the door for new tailored hybrid 
materials with specific transport, magnetic and/or superconducting properties. Albeit 
various such layered MO combinations are realized, heterostructures 
build up by interlacing band-insulating with Mott-insulating compounds are particularly 
appealing. As they raise questions about the electronic states 
resulting from conceptually quite different limits, this structural setup challenges 
the existing modern first-principles approaches to electronic structure.

Coherent superlattices (SLs) of the strongly correlated LaTiO$_3$ (LTO) Mott insulator 
with the SrTiO$_3$ (STO) band insulator belong to the most prominent examples of these 
structured materials.~\cite{oht02} Not only displays the LTO/STO interface 
metallicity, also magnetic correlations are vital due to the antiferromagnetic ordering
of LTO~\cite{mei99} below $T_{\rm N}$=146K. Moreover 2D superconductivity was 
revealved.~\cite{bis10} Aside from possible polarization mechanisms~\cite{nak06} and 
suppressing Mott correlations, the nominal Ti$^{3+}$ valence in LTO allows for doping 
the band-insulating STO side, giving rise to metallic transport. Yet in reality 
strong Coulomb interactions among the electrons complicate this simplistic picture. 
Photoemission experiments~\cite{tak06} indeed reveal strong correlation signatures, 
i.e. lower Hubbard band and quasiparticle (QP) peak, in the valence spectrum.
The interface conductivity~\cite{jan11} as well as the optical response~\cite{seo07} is 
identified to dependent on electronic correlations. Numerous theoretical studies on the 
interface electronic structure exist. There are inital tailored Hubbard-model 
considerations from unrestricted Hartree-Fock,~\cite{oka04} several first-principles 
investigation based on Kohn-Sham density functional theory 
(DFT),~\cite{pop05,ham06,lar08} DFT+U studies~\cite{oka06,pen07} as well as many-body 
approaches based on the Lanczos-method,~\cite{kan06} slave bosons~\cite{rue07} and 
dynamical mean-field theory (DMFT).~\cite{oka04_3,ish08}

Contrary to former studies, this work treats the effective single-particle 
character of the materials chemistry on an equal footing with 
many-body effects from a local perspective. We performed charge self-consistent
DFT+DMFT~\cite{sav01,pou07} computations for selected LTO/STO SLs that allow for detailed 
examinations of the subtle interplay between realistic interface effects and  
multi-orbital electronic correlations at room temperature.
\begin{figure}[t]
\begin{center}
\includegraphics*[height=4cm]{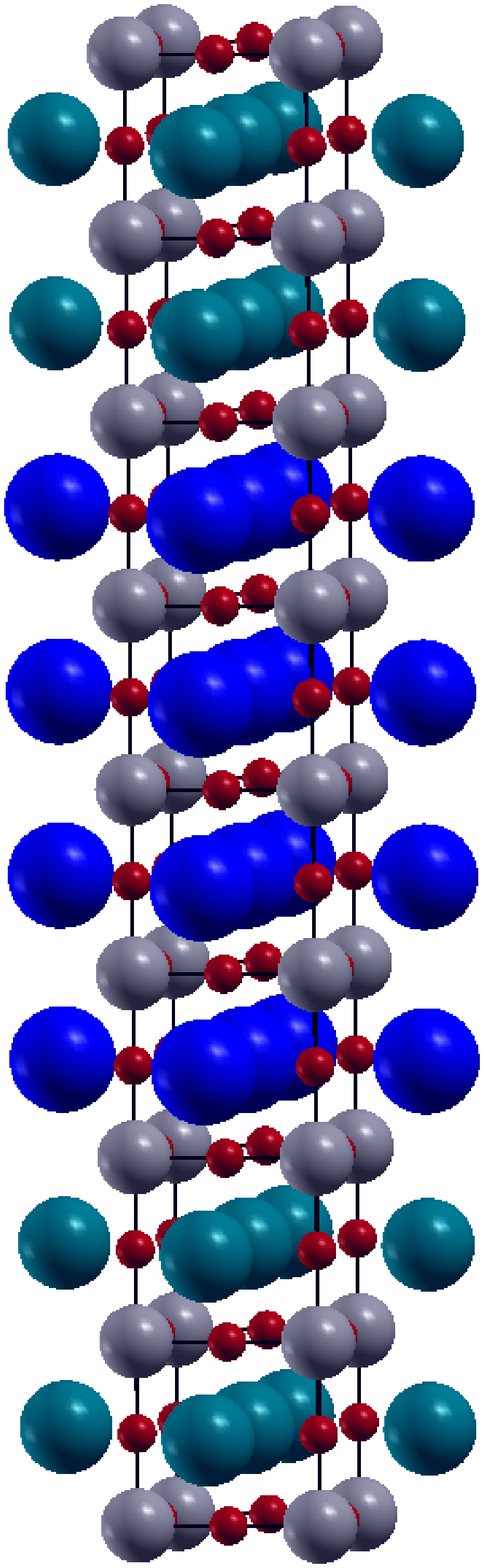}
\includegraphics*[height=4cm]{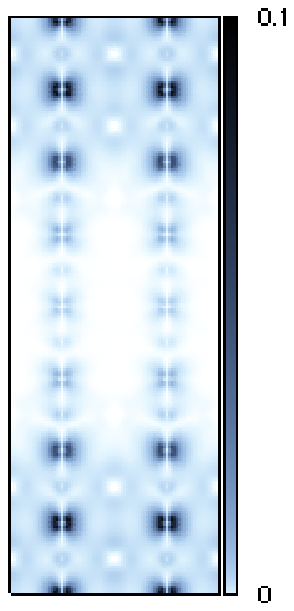}\hspace*{0.1cm}
\hspace*{-0.4cm}\includegraphics*[height=4cm]{ldabands.eps}
\end{center}
\caption{(color online) Left: (4,4)$\times$2 LaTiO$_3$/SrTiO$_3$ ideal tetragonal unit 
cell with La (big green/grey), Sr (big blue/dark), Ti (grey/light grey) and 
O (small red/dark) ions as well as the charge density from $t_{2g}$-like 
LDA bands below $\varepsilon_{\rm F}$, holding 8$\,{\rm e}^-$. Right: corresponding 
LDA bandstructure and DOS for ideal-tetragonal and orthorhombic-relaxed 
case.}\label{fig:ldabands}
\end{figure}
It is revealed that structural relaxations and electronic correlations ally in driving
an enlarged Ti$(3d_{xy})$ orbital polarization in real space {\sl and} in the low-energy 
spectrum. Fostered by enlarged lateral coherency effects, a prominent renormalized
$d_{xy}$ QP peak resides close to the Fermi level.
Moreover investigated single-layer LTO/STO architectures remain metallic troughout the
SLs.
\begin{figure*}[ht]
\begin{center}
(a)\includegraphics*[height=5.7cm]{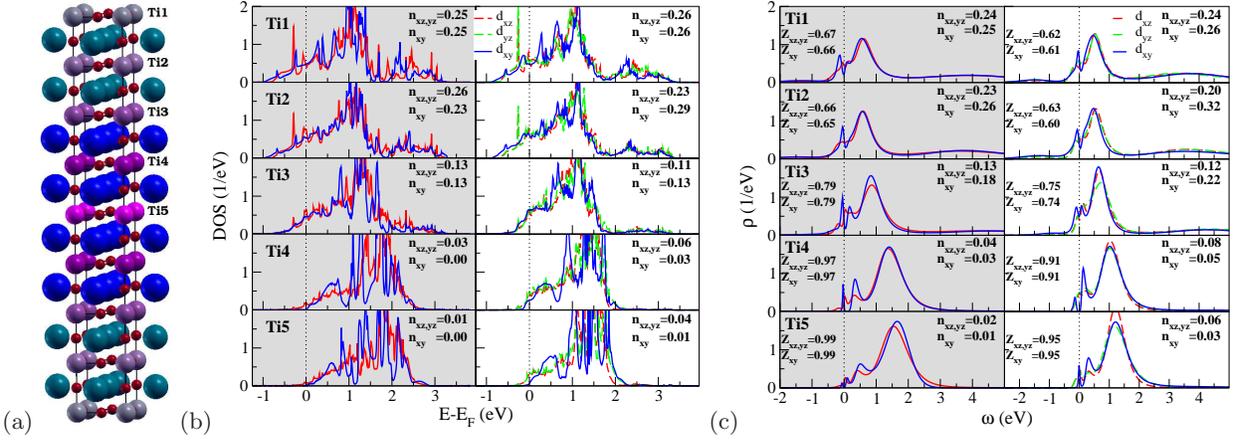}
(b)\includegraphics*[height=5.7cm]{DOS-tmo.eps}
(c)\includegraphics*[height=5.7cm]{LSPEC-tmo.eps}
\end{center}
\caption{(color online) (a) Inequivalent Ti ions for (4,4)$\times$2. 
(b) Local Ti1-5 LDA DOS for unrelaxed (grey background (bg)) 
and structurally relaxed (white bg) cases. (c) Local DFT+DMFT spectral function 
for Ti1-5 with same color coding. Though $d_{xz,yz}$ are different by symmetry with
relaxations, $Z$ and $n$ differ only marginally, hence averaged values
are shown.}\label{fig:tidos}
\end{figure*}

The theoretical approach (see Refs.~\onlinecite{ama08,gri12} for details) 
builds up on the combination of a mixed-basis pseudopotential framework~\cite{mbpp_code} 
with a hybridization-expansion continuous-time quantum Monte Carlo 
solver~\cite{rub05,wer06,triqs_code} for the DMFT impurity problem.
To include important structural relaxations~\cite{oka06,ham06,lar08} in a general
scope, allowing for layer-distance variation {\sl and} tilting of the TiO$_6$ octahedra, 
we constructed SLs in an $(n$,$m)$$\times$2 setup, where $n,m$ denote the numbers of LaO, 
SrO layers in the unit cell with two inplane Ti ions, respectively 
(see Fig.~\ref{fig:ldabands}). 
The two Ti ions are assumed equivalent by symmetry in each layer, so possible 
lateral orderings~\cite{pen07} are suppressed. For all discussed SLs
the lateral lattice constant was set to the STO value $a$=3.905\AA$\,$ and a ratio
$c/a$=0.99 was identified reasonable.

Figure~\ref{fig:ldabands} shows the LDA bandstructure for a (4,4)$\times$2 superlattice 
(80-atom unit cell) along with the density of states (DOS). The occupied bands just 
below the Fermi level $\varepsilon_{\rm F}$ with dominant Ti($t_{2g}$) character 
accommodate the eight additional electrons from the La$^{3+}$ ions in the unit cell.
Structural relaxations, relevant also in the STO part, enlarge the gap between the 
O$(2p)$-derived bands deep in energy and the latter $t_{2g}$ bands as well as increase 
the pseudogap between $t_{2g}$
and $e_g$-like states high in energy. A DOS maximum right at $\varepsilon_{\rm F}$ 
exists for both structural cases, but there is a gain of 36 meV/atom in the LDA total 
energy upon relaxation. The real-space distribution of the occupied $t_{2g}$ valence 
charge density in Fig.~\ref{fig:ldabands} elucidates the $t_{2g}$ doping in the STO part.
This charge transfer is even strengthened in the relaxed orthorhombic structure. 

The (4,4)$\times$2 unit cell contains five symmetry inequivalent Ti ions, denoted here
Ti1-5, which local $t_{2g}$ DOS from projected local orbitals~\cite{ama08} is displayed 
in Fig.~\ref{fig:tidos}b. From the mid LTO part (Ti1) to the mid STO part 
(Ti5) the $t_{2g}$ filling is decreasing. The effective bandwidth shrinks from 
$\sim$4 eV down to $\sim$2.5 eV, with minor smaller size in the relaxed
structure. Though overall rather balanced, the $d_{xy}$ occupation is somewhat 
increased by structural relaxation. In the latter case, the total number of $t_{2g}$ 
electrons is higher, i.e. the doping of these states is more efficient. 

To capture the effect of many-body correlations an effective three-orbital Hubbard 
Hamiltonian ${\cal H}$ with fully rotational invariant interaction terms, i.e.
\begin{eqnarray}
{\cal H}&=& 
U\sum_{m} n_{m\uparrow}n_{m\downarrow}+\frac{1}{2}\sum \limits _{m\ne m',\sigma}
\Big\{U' \, n_{m \sigma} n_{m' \bar \sigma}+\nonumber\\
&&+\,U'' \,n_{m \sigma}n_{m' \sigma}+
\,J\left(c^\dagger_{m \sigma} c^\dagger_{m' \bar\sigma} 
c^{\hfill}_{m \bar \sigma} c^{\hfill}_{m' \sigma}+\right.\\   
&&\left.+\,c^\dagger_{m \sigma} c^\dagger_{m \bar \sigma}  
 c^{\hfill}_{m' \bar \sigma} 
c^{\hfill}_{m' \sigma}\right)\Big\}\;,\nonumber      
\label{eq:hubbardham}          
\end{eqnarray} 
is applied at each individual Ti site $i$. Its parametrized by the 
adequate~\cite{miz95,pav04} Coulomb integral $U$=5eV and the Hund's exchange $J$=0.7eV 
with $U'$=$U$$-$$2J$ and $U''$=$U$$-$$3J$. Including symmetry, this leads to five 
inequivalent single-site impurity problems embedded in the full charge self-consistent
DFT+DMFT calculations for the (4,4)$\times$2 unit cell. Charge self-consistency is a vital
methodological ingredient because of the subtle electron transfers  
(cf Fig.~\ref{fig:ldabands}). For the projected-local-orbital 
construction of the correlated subspace a number of 80 Kohn-Sham bands starting from the 
bottom of the $t_{2g}$-like manifold was used. The double-counting correction 
applied to each Ti impurity self-energy $\Sigma_i$ amounts to an site-averaged 
fully-localized~\cite{sol94} term. All DFT+DMFT computations were performed at
$T$=290K.~\footnote{For the analytical continuation of the QMC data we used the maximum
entropy method, which we cross-checked with Pad{\'e} approximants}

In Fig.~\ref{fig:tidos}c the resulting local $t_{2g}$ spectral functions are plotted. 
Compared to LDA the total filling increases once more with correlations. 
Close to the interface especially the $d_{xy}$ 
orbital character gains further occupation with correlations. Moreover the $d_{xy}$ 
weight near $\varepsilon_{\rm F}$ is now pronounced compared to $d_{xz,yz}$, 
resulting in a dominant $d_{xy}$ QP peak below the Fermi level. Coherent
transport is thus $d_{xy}$ dominated. From the incoherent high-energy part, the 
correlation strength is larger in the structurally relaxed orthorhombic SLs. Right
at the interface (Ti3 ion) the $d_{xy}$ local spectral part is susceptible to
(pseudo)gapping. The site- and orbital-resolved QP weight 
$Z_{im}$=$(1$$-$$\partial\Sigma_{im}/\partial\omega)^{-1}$ varies significantly accross 
the interface, revealing somewhat stronger mass renormalization for $d_{xy}$.
\begin{figure}[t]
\begin{center}
\includegraphics*[width=8.5cm]{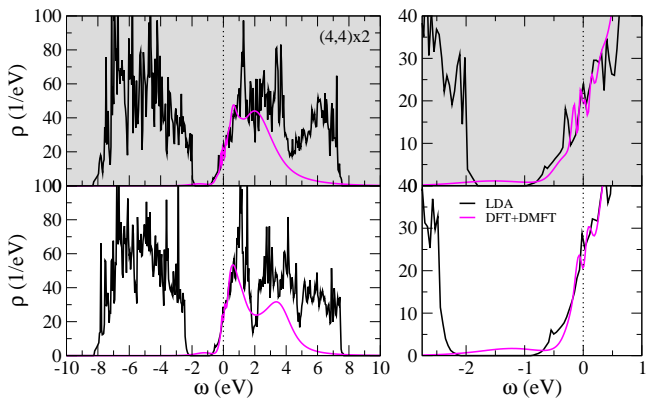}
\end{center}
\caption{(color online) Total $t_{2g}$-like DFT+DMFT spectral function  
compared to the LDA DOS for (4,4)$\times$2. 
Top: tetragonal unrelaxed (grey bg), bottom: relaxed orthorhombic (white bg). 
Right part shows a blow up close to $\varepsilon_{\rm F}$.
}\label{fig:totspec}
\end{figure}
\begin{figure}[b]
\begin{center}
\includegraphics*[width=8.5cm]{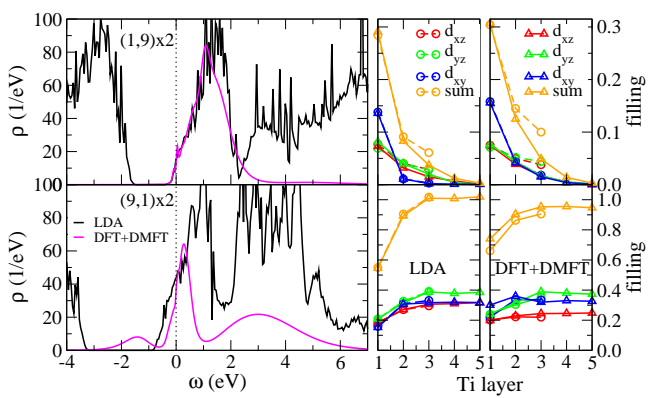}
\end{center}
\caption{(color online) Left: total $t_{2g}$-like DFT+DMFT spectral function compared 
to the LDA DOS for (1,9)$\times$2 (top) and (9,1)$\times$2. Right:
Ti($t_{2g}$) occupations with distance from the LTO/STO layer within LDA (left part)
and DFT+DMFT (right part). Circles mark results for the 4-layer host, triangles for the
9-layer host.}\label{fig:1layertot}
\end{figure}
The complete DFT+DMFT $t_{2g}$-like spectral function in Bloch space carries the 
dominance of the $d_{xy}$ close to the Fermi level (see Fig.~\ref{fig:totspec}). While 
in the unrelaxed tetragonal case a maximum at $\varepsilon_{\rm F}$ remains as in 
LDA, a minimum occurs when including structural relaxations. Increased 
spectral-weight transfer towards the lower Hubbard band takes place in the latter. This
Hubbard peak shifts closer to $-1$ eV with relaxations, in good agreement 
with photoemission.~\cite{tak06}

In order to obtain deeper insight in the relevance of correlation effects, 
lets turn now to the limiting case of a single LaO(SrO) layer within an STO(LTO) host.
Within our superlattice approach two different unit cells were chosen, respectively, 
incorporating 4 and 9 host layers. The larger structure amounts to a 100-atom unit cell. 
While the 4-layer-host case has 5 Ti layers inbetween the single layers and 3 
inequivalent Ti ions, the 9-layer-host structure has 10 Ti layers and 5 inequivalent 
ones, denoted Ti1-5 with increasing distance from the threaded single layer. 
Hence the smaller(larger) structure has an odd(even) number of Ti layers. 
In the following only the structurally relaxed orthorhombic cases are discussed. 

Figure~\ref{fig:1layertot} displays the spectral comparison between LDA and charge 
self-consistent DFT+DMFT applied to the (1,9)$\times$2 as well as the (9,1)$\times$2 
structure. The correlated spectral function of (1,9)$\times$2 with one LaO layer in 
STO is metallic in agreement with optics,~\cite{seo07} showing a smaller QP peak below 
$\varepsilon_{\rm F}$ and a more prominent one just above. While at 
low energy the LDA difference between both structural types amounts mainly to a 
Fermi-level shift (1 vs. 9$\,{\rm e}^-$ below $\varepsilon_{\rm F}$), 
DFT+DMFT signals the increased correlations for (9,1)$\times$2 via substantial 
spectral-weight transfer to Hubbard bands.
\begin{figure}[b]
\begin{center}
\includegraphics*[height=4.45cm]{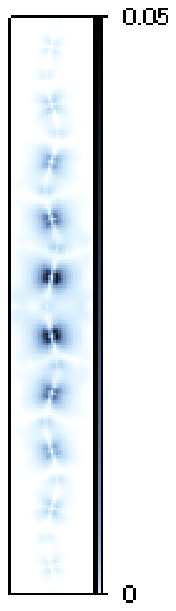}
\hspace*{-0.55cm}\includegraphics*[height=4.4cm]{single-spec.eps}
\hspace*{-0.15cm}\includegraphics*[height=4.45cm]{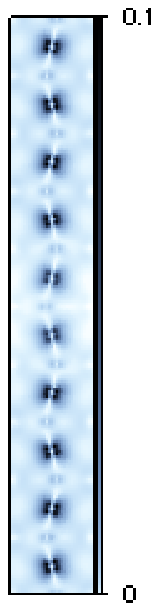}
\end{center}
\caption{(color online) Left: LDA charge density from occupied $t_{2g}$-like  
bands for a Ti column along the $c$ axis of (1,9)$\times$2. 
Right: same for (9,1)$\times$2. 
Middle: LDA $t_{2g}$-like DOS (left) and DFT+DMFT spectrum (right) for 
Ti1, Ti5 (see text).}\label{fig:single-spec}
\end{figure}
No insulating state, also not site-selective (see below), is obtained for 
(9,1)$\times$2, which thus could be viewed as a doped Mott-insulator. Note that also
bulk Sr$_{1-x}$La$_x$TiO$_3$ is insulating only above $x_c$$\sim$0.95.~\cite{tok93}
Experiments on confined STO in a larger GdTiO$_3$ host yet show the principle
chance for reaching an insulating interface~\cite{moe12} and the finding stimulated
modeling ideas based on Hubbard-ladder physics.~\cite{che13}

Instructive are the local $t_{2g}$-like occupations for Ti1-5.
For the single LaO layer the Ti doping in STO is still not accomplished in  
(1,4)$\times$2 far from the interface. But for (1,9)$\times$2 the Ti5 ion indeed shows 
zero $t_{2g}$ filling (see also Fig~\ref{fig:single-spec}). DFT+DMFT leads here to 
somewhat farther charge flow into STO and a stronger orbital-filling alignment within 
the $t_{2g}$ manifold away from the interface. However again the $d_{xy}$ polarization 
close to it is strengthened with correlations. The evaluated sheet carrier densities 
(in unit cm$^{-2}$) $n_{\rm sheet}^{\rm LDA}$=1.8$\times$$10^{14}$ and 
$n_{\rm sheet}^{\rm DFT+DMFT}$=2.3$\times$$10^{14}$ agree well with
the experimental value $n_{\rm sheet}^{\rm exp}$$\approx$3$\times$$10^{14}$ from 
optics.~\cite{seo07} The increased charging of the Ti1 ion in 
DFT+DMFT for the single SrO layer architecture (again favoring $d_{xy}$) is evident. In
general, whereas LDA quickly saturates here to the nominal $n$=1 $t_{2g}$ occupation,
many-body effects result in a balancing of the strong LDA occupation differences
with distance from the interface. Even for Ti4,Ti5 the hole doping is vital. A 
subtle $d_{xy}$/$d_{yz}$ filling crossover occurs near Ti2, which marks the 
competition between bulk-LTO Mott-insulating behavior and LTO/STO interface physics. 
While in the former case indeed the $d_{yz}$ orbital has dominant contribution to the 
correlated crystal-field ground state,~\cite{pav05} the driving force for $d_{xy}$ 
polarization is stronger at the interface.
\begin{figure}[t]
\begin{center}
\includegraphics*[width=7.5cm]{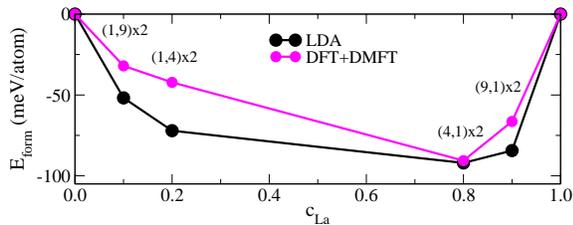}
\end{center}
\caption{(color online) Formation energy and convex hull for the embedded single-layer 
SLs.}
\label{fig:eform}
\end{figure}

Besides illustrating the real-space variation of the low-energy LDA valence charge 
density for (9,1)$\times$2 and (1,9)$\times$2, in Fig.~\ref{fig:single-spec} the local 
$t_{2g}$-like spectral properties are visualized for the Ti ions closest (Ti1) and 
farthest from the single LaO(SrO) layer. The QP structure for Ti1 in (1,9)$\times$2 is
more subtle than for the interface ions in (4,4)$\times$2, since a strong 
orbital-selectivity towards $d_{xy}$ has not yet manifested in the single-layer set up.
Still a shifting of the large $d_{xz,yz}$ DOS below $\varepsilon_{\rm F}$ towards the
unoccupied part is visible. The $t_{2g}$-like orbital behave more balanced in 
(1,9)$\times$2 with significant Hubbard-band weight for Ti5. However a standard 
doped-Mott-insulating picturing of the Ti5 multiorbital spectral function seems not 
applicable, i.e. the QP weight remains sizeable.

Finally we want to comment on the energetics of the relaxed embedded single-layer SLs 
within LDA and  DFT+DMFT.~\cite{gri12} To this we compute the formation energy, defined 
here as
\begin{equation}
E^{\rm form}_{n,m}=E^{\rm tot}_{n,m}-c_{\rm La}E^{\rm tot}_{\rm bulk-LTO}
-(1-c_{\rm La})E^{\rm tot}_{\rm bulk-STO}\;
\label{eq:eform}
\end{equation} 
where $E^{\rm tot}$ denotes the total energy per atom and $c_{\rm La}$=$n/(n$$+$$m)$. Be
aware of the nontrivial character, since in the correlated case not only 
$E^{\rm form}_{n,m}$ includes many-body corrections but also $E^{\rm tot}_{\rm LTO,STO}$. 
As the projected local orbitals are here derived from $t_{2g}$-like bands, however
that correction vanishes for bulk-STO having those unoccupied. Yet bulk-LTO is
a Mott insulator in DFT+DMFT,~\cite{pav04} which is verified within the charge 
self-consistent scheme for the here given many-body Hamiltonian (1),
chosen Coulomb parameters and double counting. Within our double-counting scheme
the bulk-LTO correlated Mott state enters eq.~(\ref{eq:eform}) with a much lower total
energy. Figure~\ref{fig:eform} shows the variation of the formation energy based on the 
four single-layer $(n$,$m)$$\times$2 structures.
Already in LDA an asymmetry of the basic convex hull towards the LTO side exists, 
which is enforced with correlations. Given the delicacy of $E^{\rm form}$, the
values within both theoretical schemes are rather similar, which especially speaks for 
the novel DFT+DMFT approach to such subtle energetic quantities. The observed trend of 
generally reduced $E^{\rm form}_{n,m}$ (though evaluated at finite $T$) may correct 
frequent LDA overestimations.

In summary, the correlated electronic structure of realistic LTO/STO SLs has been
studied witin
charge self-consistent DFT+DMFT beyond static mean-field. Many-body effects and 
structural relaxations are revealed to ally in driving orbital-selective behavior 
towards dominant Ti$(3d_{xy})$ filling {\sl and} QP behavior. In addition, significant 
spectral-weight intensity just above the Fermi level renders intricate susceptibility 
to applied field possible. Single-layer architectures of these systems result in global, 
but still intriguing, metallic behavior for the studied embeddings. In general, the 
recent advances put the DFT+DMFT formalism in position for challenging materials 
investigations, including engineered systems. A further methodological step 
will be the sound computation of correlation-influenced structural relaxations. Future 
DFT+DMFT work on MO heterostructures such as the study of possible ordering 
instabilities, of different interface geometries and of spin-orbit effects is envisaged.

\begin{acknowledgments}
Calculations were performed on the JUROPA cluster of the J\"ulich Supercomputing Centre 
(JSC). This research was supported by the DFG-FOR1346 project.
\end{acknowledgments}

\bibliographystyle{apsrev4-1}
\bibliography{bibextra}

\end{document}